\documentclass{article}

\PassOptionsToPackage{numbers, compress}{natbib}

\usepackage[final]{hraim_2024}

\usepackage[utf8]{inputenc} 
\usepackage[T1]{fontenc}    
\usepackage{hyperref}       
\usepackage{url}            
\usepackage{booktabs}       
\usepackage{amsfonts}       
\usepackage{nicefrac}       
\usepackage{microtype}      
\usepackage{xcolor}         
\usepackage{graphicx}
\usepackage{lipsum}
\usepackage{natbib}
\usepackage{amsmath}
\usepackage{amssymb}
\usepackage{floatrow}
\newfloatcommand{capbtabbox}{table}[][2\FBwidth]
\usepackage{blindtext}
\usepackage{multirow}
\usepackage{graphicx}

\title{Balancing Power and Ethics: A Framework for Addressing Human Rights Concerns in Military AI}

\author{%
Mst Rafia Islam\textsuperscript{1}, Azmine Toushik Wasi\textsuperscript{2}\\
  \textsuperscript{1}Independent University, Bangladesh 
  \textsuperscript{2}Shahjalal University of Science and Technology, Bangladesh\\
  \texttt{2030391@iub.edu.bd, azmine32@student.sust.edu} \\
  Both authors contributed equally to the project
}

\begin{document}

\maketitle
\vspace{-10mm}
\section{Introduction}
\vspace{-4mm}
AI has advanced significantly in recent years, resulting in a wide range of applications in both the civilian and military sectors. The military, which is continually seeking innovations for more effective, faster, and stronger technology or weapons, sees AI as a perfect answer to address these needs \cite{Fischer2022,GrandClement2023}. AI technologies offer many advantages, such as increased operational efficiency, precision targeting, and reduced human casualties \cite{Rashid2023,Murray2024arfAFRAF}. However, these advancements come with profound ethical and legal concerns, particularly regarding potential human rights violations \cite{Bradley2020,Murray2024arfAFRAF,AU-AI_Act}.

\vspace{-1mm}

Use of AI in military applications raises serious concerns \cite{Marwala2023,Adam2024,GrandClement2023}. Autonomous weapons systems, capable of making life-or-death decisions without human intervention, threaten the right to life and violate international humanitarian law. While AI-enhanced surveillance systems improve data collection, they can lead to widespread privacy violations and unnecessary monitoring of individuals \cite{GrandClement2023}. Additionally, inherent biases in AI systems may result in discriminatory practices, exacerbating existing disparities and violating principles of equality, non-discrimination, and human rights \cite{Bradley2020,Murray2024arfAFRAF}.

\vspace{-1mm}

Motivated by these concerns, in this work, we propose a novel three stage framework to address human rights issues in the design, deployment, and use of military AI, with each aspect consisting of multiple components.
Our work aims to propose a comprehensive framework for evaluating human rights violations by military AI, addressing both technical and statutory aspects. By examining the ethical implications and regulatory challenges, this framework seeks to provide a balanced approach to harnessing the benefits of AI in military operations while safeguarding human rights.

\vspace{-4mm}
\section{Framework}
\begin{figure}[t]
    \centering
    \includegraphics[width=\linewidth]{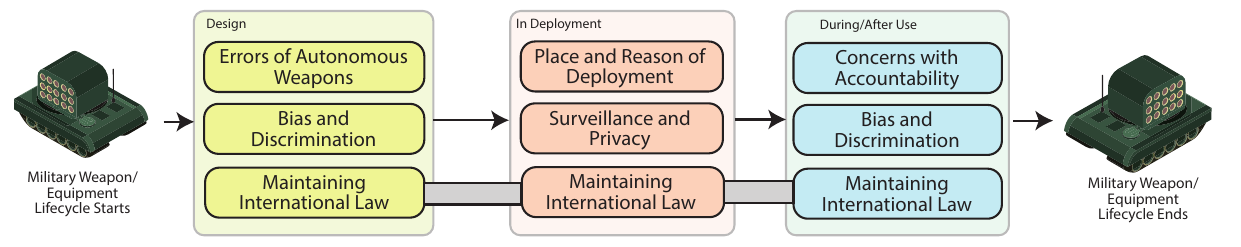}
        \vspace{-6mm}
    \caption{Our framework for addressing human rights concerns in military AI}
    \label{fig:WHAT}
\end{figure}

\vspace{-4mm}
As shown in Figure \ref{fig:WHAT}, this framework addresses human rights concerns in the Military AI lifecycle across three key phases: \textbf{Design, Deployment, and Use}, with each phase having several components. Below, we discuss each phase and its components in detail.

\vspace{-3mm}
\subsection{Design Phase}
\vspace{-2mm}
Design phase involves creating military AI systems, focusing on algorithms and functionalities. It sets the foundation for AI behavior, ensuring it operates correctly, without bias and discrimination.

\textbf{Targeting Errors of Autonomous Weapons.} \quad AI-powered autonomous weapons systems can make life-and-death decisions without human intervention \cite{Christie2023}. AI systems can make errors in identifying and targeting individuals, leading to unintended casualties and injuries among civilians \cite{Asaro2012}. Implementing ethical guidelines in the design and development of military AI systems can ensure compliance with human rights standards. 
Moreover, incorporating human oversight in AI decision-making processes ensures accountability and ethical compliance. Human-in-the-loop (HITL) \cite{Mosqueira-Rey2023-zv} systems allow human operators to intervene and override AI decisions when necessary.

\vspace{-1mm}

\textbf{Bias and Discrimination.} \quad AI systems can perpetuate or even exacerbate existing biases, leading to discriminatory practices in targeting or profiling based on race, ethnicity, or other characteristics \cite{Gordon2019}. By developing algorithms that detect and mitigate biases in AI systems can prevent discriminatory outcomes. This not only enhances the fairness of AI applications but also helps in building public trust and ensuring compliance with ethical standards.

\vspace{-4mm}
\subsection{In Deployment Phase}
\vspace{-3mm}
In Deployment phase involves integrating military AI systems into specific environments, focusing on their usage while emphasizing ethical considerations and legal compliance.

\textbf{Surveillance and Privacy.} \quad   AI-driven surveillance systems can infringe on individuals’ privacy rights, leading to unauthorized monitoring and data collection \cite{feldstein2019global}. Implementing strong data privacy and security measures to protect sensitive information collected and processed by military AI systems. This includes encryption, access controls, and regular security assessments. 

\textbf{Deployment of the weapon.} \quad  To ensure maximum efficacy while minimizing collateral damage \cite{Stein2007-rm}, the automated weapon system must be deliberately positioned in a location that maximizes operational range and precision. In order to prevent unintended harm, this entails a detailed investigation of the terrain, potential targets, and civilian areas \cite{Crootof2014TheKR}. Continuous monitoring and changes are required to ensure its efficacy and safety. Additionally, real-time data analysis and feedback loops should be integrated to adapt to dynamic battlefield conditions, ensuring that the system remains responsive and minimizes risks to non-combatants \cite{Asaro2012}.

\vspace{-4mm}
\subsection{During/After Use Phase}
\vspace{-3mm}
During/After Use phase encompasses the actual operation of military AI systems during conflicts or missions. It assesses the AI's performance in real-time, monitoring for accountability and compliance with human rights and legal standards.

\textbf{Concerns with Accountability.} \quad   Transparency of AI decision-making processes complicates accountability, making it difficult to attribute responsibility for human rights violations \cite{Novelli2024-jz}.  When AI systems make decisions, it creates a gap in justice and responsibility, as it becomes challenging to hold any individual or entity accountable for wrongful actions \cite{Pasquale2015}. By ensuring AI systems are transparent and their decision-making processes are explainable, accountability issues can be resolved. Ethical guidelines, such as the Asilomar AI Principles \cite{Asilomar2017}, emphasize the importance of transparency, accountability, and human oversight in the development and deployment of military AI. Adhering to these principles can help mitigate the violations of human rights. 

\textbf{Bias and Discrimination.} \quad AI systems can perpetuate or worsen existing biases, leading to discriminatory practices in targeting or profiling based on race, ethnicity, or other characteristics \cite{feldstein2019global}. To mitigate these biases post-deployment, continuous monitoring and auditing of AI systems are essential. Implementing feedback loops to identify and correct biased outcomes can ensure fairness \cite{Raji2020-gk}. This approach addresses biases and helps build public trust in AI technologies by demonstrating a commitment to ethical standards and accountability.

\vspace{-4mm}
\subsection{Violation of International Humanitarian Law}
\vspace{-3mm}
Deployment of AI in military operations poses serious risks to International Humanitarian Law (IHL), which aims to protect non-combatants. Autonomous weapons can make decisions without human oversight, leading to indiscriminate attacks that violate the principle of distinction \cite{Asaro2012}. AI targeting decisions may also compromise proportionality, as these systems might not effectively balance military advantage with civilian harm \cite{Crootof2014TheKR}. The lack of transparency and accountability in AI decision-making complicates compliance with IHL, hindering the investigation and attribution of unlawful actions. Adherence to IHL requires strong ethical guidelines, oversight mechanisms, and international cooperation to prevent violations and uphold the laws of war.

\vspace{-4mm}
\section{Challenges and Future Work}
\vspace{-4mm}
Several challenges may hinder the proper implementation of our framework. There is no global agreement on the ethical use of military AI. Groups like the Campaign to Stop Killer Robots push for a ban, but leading AI nations resist binding commitments. Accountability is also complex, as it's hard to determine who is responsible when AI systems cause harm—creators, commanders, or the AI itself. Additionally, AI evolves quickly, outpacing legal systems and creating oversight gaps that could be exploited. Future research should focus on establishing international agreements and fostering collaboration between engineers, ethicists, and legal experts to keep pace with AI advancements.

\small{
\bibliography{hraim_2024}
}




\end{document}